\documentclass[aps,prmaterials,twocolumn,a4,nofootinbib,showkeys]{revtex4-2}

\usepackage{float}
\usepackage[T1]{fontenc}
\usepackage{framed}
\usepackage{graphicx}

\newcommand{\ads}{{\mbox{\scriptsize ads}}}
\newcommand{\MgX}{{\mbox{\scriptsize MgX}}}
\newcommand{\alloy}{{\mbox{\scriptsize X}}}
\newcommand{\nn}{{\mbox{\scriptsize nn}}}

\newcommand{\tot}{{\mbox{\scriptsize tot}}}

\newcommand{\wat}{{\mbox{\scriptsize water}}}
\newcommand{\vac}{{\mbox{\scriptsize vac}}}

\newcommand{\mol}{{\mbox{\scriptsize mol}}}
\newcommand{\surf}{{\mbox{\scriptsize surf}}}

\newcommand{\full}{{\mbox{\scriptsize full}}}
\newcommand{\sol}{{\mbox{\scriptsize sol}}}
\newcommand{\vg}{{\mbox{\scriptsize vac-geo}}}
\makeatletter

\begin{document}


\title{A density functional theory study of amino acids on Mg and Mg-based alloys}

\author{John Bolin, Amanda Goold, Olof Hildeberg, Alva Limb\"ack and Elsebeth Schr{\"o}der}
\email{Corresponding author: E. Schr\"oder, schroder@chalmers.se}%
\affiliation{Microtechnology and Nanoscience, MC2,
Chalmers University of Technology, SE-41296, Gothenburg, Sweden}

\date{January 12, 2026}

\begin{abstract}
Magnesium (Mg) has mechanical properties similar to bone tissue, and Mg ions take part in the metabolism. 
This makes Mg of interest for biocompatible degradable body implants, provided that its high corrosion rate can be inhibited.
Slightly alloying Mg and adding surface coatings can slow down the corrosion processes without significantly changing the mechanical properties. 
Use of coating molecules that are native to the body increase the likelihood of making the surface biocompatible, for example by use of amino acids. 
We here present a density functional theory (DFT) study of the adsorption on Mg(0001)
of the amino acids glycine, L-proline, and L-hydroxyproline (Hyp), the main amino acid content of collagen. 
We investigate how binding of the functional groups of Hyp are affected when Mg(0001) is slightly alloyed with zinc, lithium or aluminium, 
and we also model the immersion of the systems in a water environment to see how this affects the binding.
\end{abstract}

\keywords{
Density functional theory, Magnesium, vdW-DF, Amino acids, Biodegradable implants
}
\maketitle

\hyphenation{over-estimated mole-cules}

\section{Introduction}
Magnesium (Mg) alloys are attractive materials for use in biodegradable implants because of their mechanical and biocompatible properties. 
The Mg alloys are light weight, have a density close to that of bone, show a
high strength-to-weight ratio, and have a stiffness more similar to that of bone than, e.g., titanium. 
Mg is an essential element in the human body, absorbed through 
the metabolism \cite{LiZheng13,rahman20,agarwal16}. 
Pure or nearly pure Mg degrades rapidly by corrosion in an uncontrolled 
manner, leading both to a very fast breakdown of the materials and to the production of excessive amounts of hydrogen gas (H$_2$) from the 
corrosion process, more than what can be handled by the body \cite{noviana16,amara23}. 
Even at low concentration, alloying can significantly alter surface properties, helping control the corrosion,
although not necessarily damping it sufficiently to keep the implant stability at the time scale needed in temporary implants. 
Coating the surface may further control the corrosion rate, and at the same time improve the implant interface with body tissues.
The goal is to keep the corrosion rate sufficiently low and the structure sufficiently stable such that
the implant keeps its mechanical integrity until  
the bone healing process is completed, often on the time scale of about a year \cite{amara23}. 

From a mechanical point of view, coating is possible with a long range of different materials. 
However, avoiding toxic materials and keeping the implant biocompatible sets limits on the material that can be used. 
As discussed in, for example Ref.\ \cite{ashassi19}, amino acids have several positive 
properties acting as part of coating layers in the implants: they are environmentally safe, 
non-toxic, already present in living organisms, and cheap and easy to prepare experimentally. 
From a calculational point of view, an added benefit is their relatively small sizes. 

We here investigate by first-principles theoretical calculations how surfaces of Mg and sparsely alloyed Mg 
interact with some of the most abundant amino acids of the body: glycine (Gly), L-proline (Pro), and L-hydroxyproline (Hyp). 
The purpose is two-fold: We want to gain a better understanding of the amino acids as potential (biomimetic) coatings 
of Mg-based implants, 
and we aim to model the interaction of collagen, consisting mainly of these selected amino acids,
with Mg alloys as the initial part of the adhesion process to biomaterials.

\begin{figure}[tb]
\centering
\includegraphics[width=0.19\columnwidth]{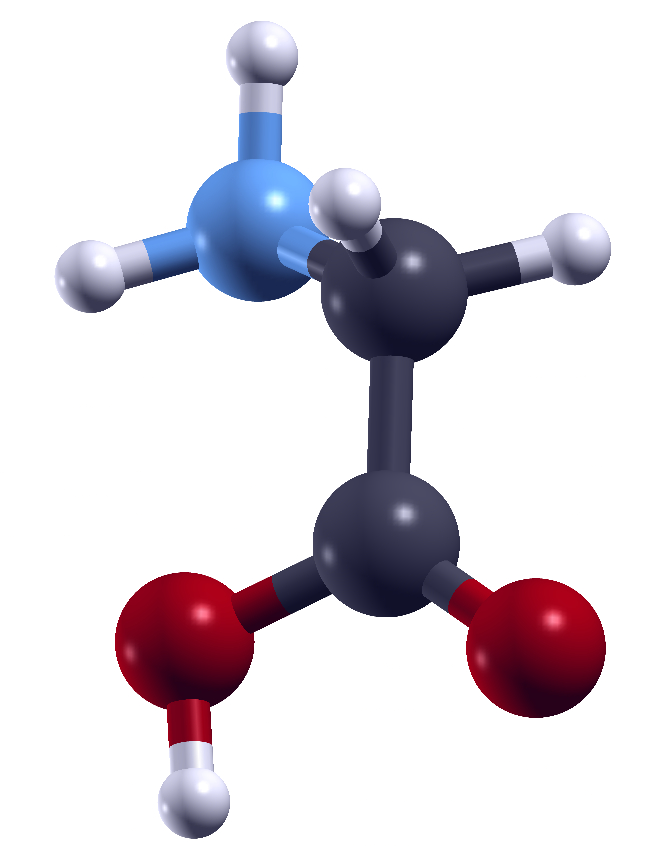}\hspace{0.08\columnwidth}
\includegraphics[width=0.19\columnwidth]{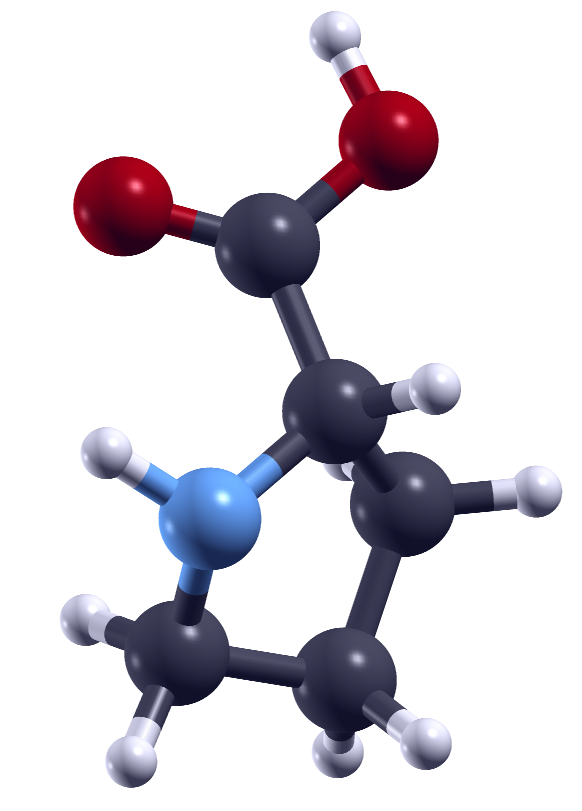}\hspace{0.08\columnwidth}
\includegraphics[width=0.19\columnwidth]{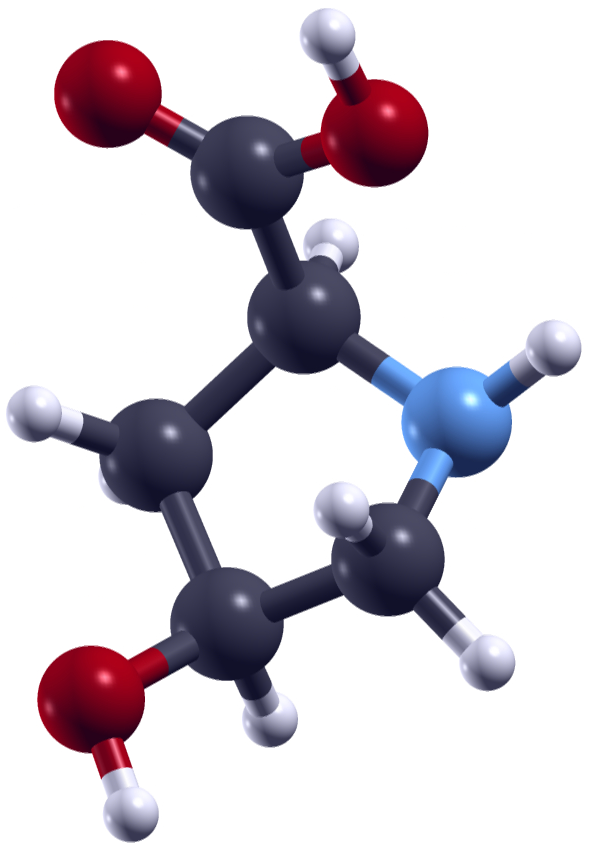}\\[0.4em]
\includegraphics[width=0.63\columnwidth]{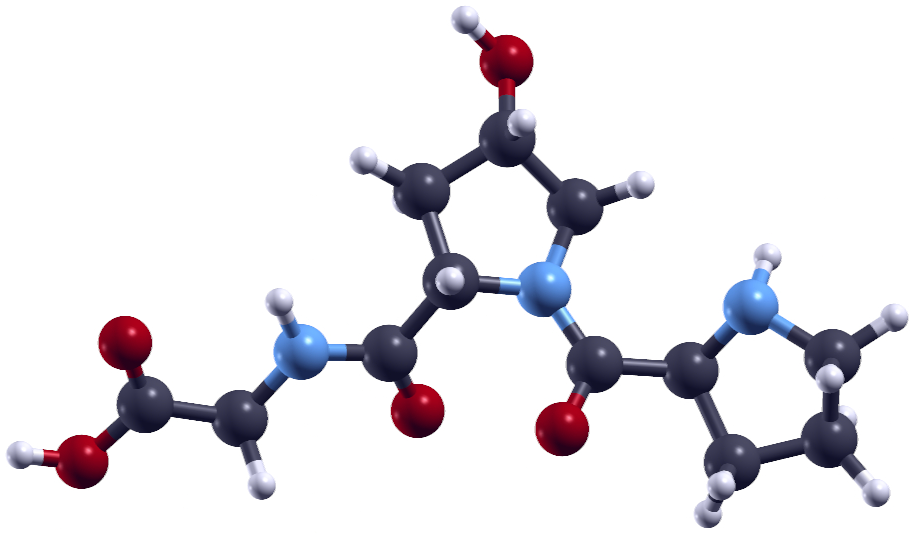}
\caption{\label{fig:molecules}\textit{Top panels:} The three amino acids studied here - glycine (Gly, C$_2$H$_5$NO$_2$), 
L-proline (Pro, C$_5$H$_9$NO$_2$), and L-hydroxyproline (Hyp, C$_5$H$_9$NO$_3$).
\textit{Bottom panel:} Gly-Hyp-Pro snippet of a strand in the collagen (Col-I) triple helix. 
Shown atoms are oxygen (red), nitrogen (blue), carbon (black), and hydrogen (white).}
\end{figure}

The text is organized as follows. Section II introduces the surfaces and molecules studied here and
Section III presents the calculation methods used.  Section IV
presents and discusses our results for
alloying of the Mg(0001) surface, the adsorption of the amino acids on these surfaces, 
and a modelling of the adsorption of collagen, based on the adsorption of a three-amino-acid snippet of a collagen strand. 
We also discuss the results and differences of adsorbing amino acids in a water environment. Finally, Section V
contains our conclusions.


\section{Material: Molecules and surfaces}

\subsection{Molecules}
Collagen is an essential protein in the intercellular matrix, and it makes up 30\% of the protein components of the
body \cite{shoulders09}. 
It is thus one of the molecules that will be present in the environment of the bone implants. 
The most common collagen, Col-I, consists of a triple helix of 
amino acid strings with a repeated sequence of amino acids Gly-X-Y, where every third of the amino acids is Gly,
and X and Y are any other amino acids, most often they are the secondary, cyclic, amino acids Pro and Hyp. 
Gly, Pro, and Hyp in total make up about half of the amino acid units 
in collagen \cite{brodsky97}. 

The atomic structures of Gly, Pro and Hyp are illustrated in Figure \ref{fig:molecules}. 
Amino acids contain the functional groups carboxyl (-COOH) and amine (-NH$_2$), or
in a few cases instead a secondary amine (-NH) with two bonds to carbon in a pyrrolidine loop, 
as is the case for the secondary amino acids Pro and Hyp.
The role of Gly in collagen is mostly to stabilize
the triple helix by hydrogen bonds from the amino group in Gly in one strand to a carbonyl group (-C=O) in carboxyl 
in one of the two adjacent strands, either by direct hydrogen bonds or by water-mediated bonds.
Gly is therefore mostly positioned closer to the centre of the collagen helix than the other amino acids.
In contrast, the pyrrolidine loops of Pro and Hyp extend out of the helix. 
Gly is thus less available to interactions of collagen with the environment, and is therefore less
relevant when studying the interaction of collagen with the implant surface.
However, from a coating point of view, where several amino acids are included in establishing coatings, 
there is no reason to not include Gly \cite{ashassi19}.

We focus this study on the amino acids Gly, Pro and Hyp, and a snippet of a collagen strand, consisting of a molecule each of Gly, Hyp and Pro.
These are natural choices for studying collagen interaction with implants. 
For coating purposes, other amino acids (or similar molecules) might also be relevant,
but the three amino acids Gly, Pro and Hyp are prototypes of biocompatible molecules and
are therefore also in focus here.

The functional groups of the molecules are, for Gly: carboxyl and amine, for Pro: carboxyl and secondary amine in the pyrrolidine loop, and Hyp: as for Pro, 
but with an additional carbonyl on the loop. 
When in the collagen strand, 
the carboxyl groups each give up a hydroxyl group (-OH) and the H in the (secondary) amine group. 
Compared to the amino acids by themselves, the three-amino-acid snippet that we study here
thus has two less hydroxyl groups and
H atoms, due to the bonds formed between Gly-Hyp and Hyp-Pro, bottom panel of Figure \ref{fig:molecules}.
The snippet is therefore expected to bind less to the surfaces than the sum of the adsorption energies of the three molecules by themselves, already before taking steric hindrance into account.

\subsection{Surfaces}
The hexagonally close packed (hcp) structure of Mg is preserved if only a small fraction
of Mg atoms are replaced by alloying atoms (doping), but the charge distribution in the surface changes,
and thus the corrosion and adsorption properties.
We here study adsorption of molecules on the clean and doped Mg(0001) surface. 
Mg(0001) is the energetically optimal surface for clean Mg and can be seen as the typical surface of Mg from
which we can learn, while other low-index surfaces such as Mg(10$\bar{1}$0) and Mg(11$\bar{2}$0) have higher surface
energies and are more reactive \cite{hagihara16,xing24}. 

The surfaces with a small alloying content are obtained by substituting single Mg atoms in the 
top atomic layer by one of the elements lithium (Li), zinc (Zn) or aluminium (Al).
All three elements are known to improve the mechanical strength and corrosion 
resistance of the resulting alloy.
Alloys with Zn may be relevant for implant materials, with Zn being an important and abundant 
trace element in the human body \cite{tapiero03}. 
Meanwhile, Al is considered neurotoxic and a possible cause of several neurological disorders \cite{elrahman03,alzheim19,agarwal16,mirza17}.
For Li the situation is less clear: In Mg-based orthopaedic implants it might pose a health risk, affecting the kidneys and the 
central nervous system, but the long-term effects are unclear \cite{bernard14}. 
However, it has also recently been suggested in degradable Mg implants to deliver local Li for treatment of  bi-polar disorder and possibly also  other disorders \cite{bath24}. 
In our present study we include Al and Li, if only to use them as comparison to Zn and to study trends.

To learn about adsorption on the alloyed Mg surfaces we focus on Hyp, because Hyp has the most functional groups of the three amino acids. 
We study the alloying atom in strategic positions relative to the functional groups of the molecule, as will be detailed in 
the Methods section.


\section{Methods}

\subsection{Density functional theory calculations}
Our calculations are based on first-principles calculations with density functional theory (DFT).
We use the plane wave code \texttt{pw.x} of the Quantum ESPRESSO suite \cite{QE,espresso,QE_2}
with our widely tested functional vdW-DF-cx \cite{dion04p246401,berland14p035412,berland14jcp,schroder17chapter}. 

We use PAW-based \cite{PAW} pseudopotentials (PPs) from the PSLibrary \cite{dalcorso14,pslibrary}, 
created with the \texttt{ld.x} code, 
except the Mg PP which was created by one of us in the earlier study \cite{schroderPP}, also by using \texttt{ld.x}.\footnote{We use our Mg PP \texttt{Mg\_pbesol\_paw\_3d.UPF} defined and tested in Ref.\ \cite{schroderPP}, and the following 
PPs from PSLibrary: \texttt{H.pbesol-kjpaw\_psl.1.0.0.UPF},\\ 
\texttt{N.pbe-n-kjpaw\_psl.1.0.0.UPF},\\ 
\texttt{C.pbesol-n-kjpaw\_psl.1.0.0.UPF}, \\
\texttt{O.pbesol-n-kjpaw\_psl.1.0.0.UPF}, \\
\texttt{Zn.pbesol-dn-kjpaw\_psl.1.0.0.UPF},\\
\texttt{Al.pbesol-nl-kjpaw\_psl.1.0.0.UPF},\\
\texttt{Li.pbesol-sl-kjpaw\_psl.1.0.0.UPF}} 
When possible, we use PBEsol-based PPs, because they generally work best with vdW-DF-cx \cite{schroderPP} and no
PP generator with the vdW-DF-cx functional is yet available.
However, the PP for N from PSLibrary
causes convergence problems in the PBEsol version and we therefore use the PBE version, for just this element. 
In all calculations we use plane wave kinetic energy cut off 50 Ry, and 400 Ry for the electron densities, determined from convergence tests.

Our calculations are carried out with periodically repeated boundary conditions. 
For Mg(0001) we model the surface with five layers of Mg atoms. 
We have earlier \cite{schroderPP} found that for accurate studies of core level shifts originating from \textit{inside\/} the Mg(0001) surface, e.g., 
from within the first five Mg layers, 
it is sufficient to use a surface slab of 9 Mg layers that are ``shaved'' off from a fully relaxed 23-layer slab.
In the present study with interactions  \textit{above\/} or in the very top layer of the Mg atoms, we argue 
that it suffices with five layers of Mg, of which 
 the bottom three layers are kept in the position they had in the 23-layer slab. 

In the plane of the surface we use $p(5\times5)$ hcp unit cells. 
In Ref.\ \cite{schroderPP} one of us determined the Mg hcp
lattice constants of the Mg PP used here, Table \ref{tab:lattconst}, thus the side lengths of our surface supercell are 
 16.0 {\AA}. 
We use 36.0 {\AA} in the direction perpendicular to the surface, which includes 25.5 {\AA} of vacuum above the surface
and at least 16.6 {\AA} of vacuum when the largest of the amino 
acid molecules is adsorbed.

In Ref.\ \cite{schroderPP} we found the Monkhorst-Pack \cite{monkhorst76p5188} k-point sampling 
for Mg hcp bulk to be highly converged at $40\times40\times24$ k-points, and that for lower accuracies 
already $10\times10\times 8$ k-points are sufficient, corresponding to $2\times2\times1$ k-points in our surface model. 
Due to the many degrees of freedom in positioning the molecules on the surfaces, we do not expect an energy accuracy better than 10 meV per adsorbed molecule, 
which means that the latter is sufficient for the present study.

\begin{figure}[tb]
\centering
\includegraphics[width=0.8\columnwidth]{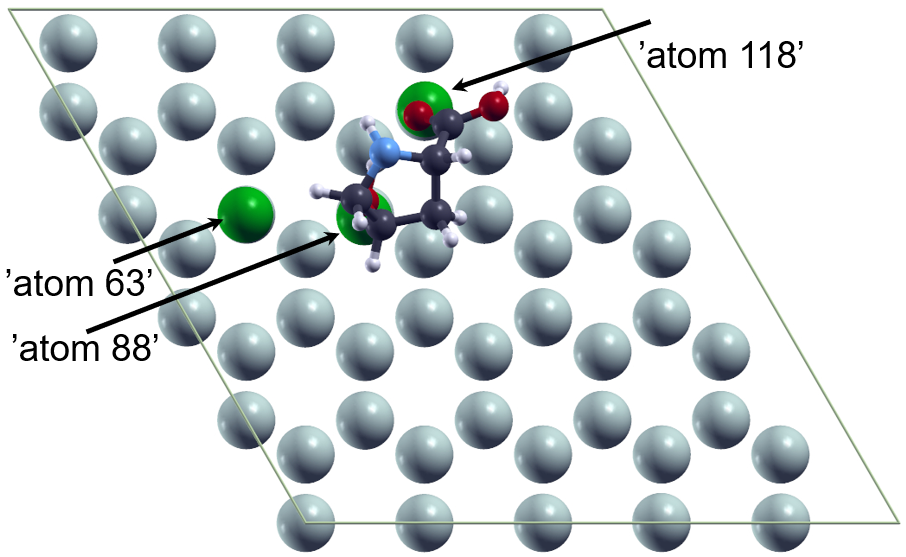}\\[1em]
\includegraphics[width=0.8\columnwidth]{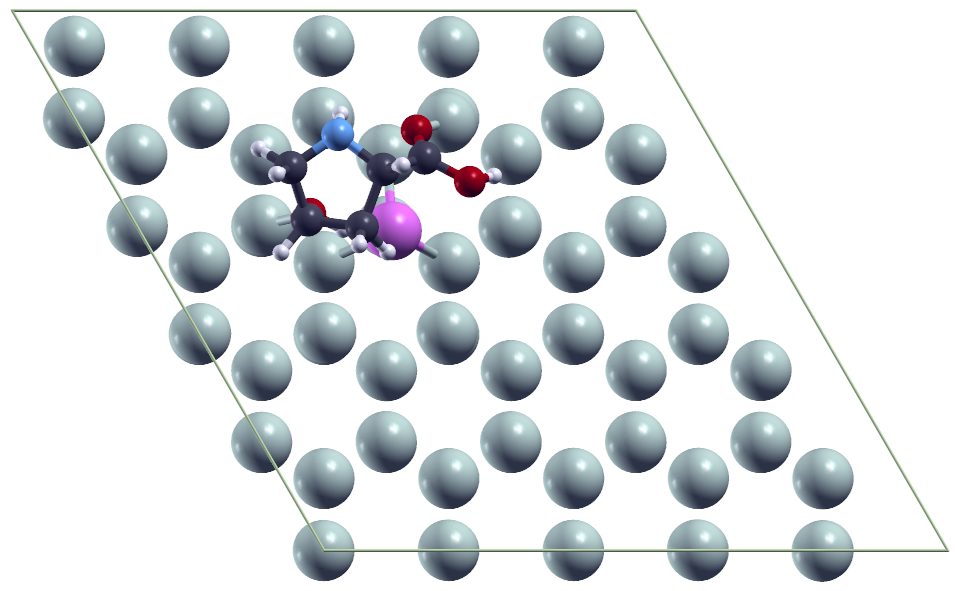}\\
\caption{\label{fig:dopingMg}\textit{Top panel:} Positions of the alloying atoms in Mg(0001) with adsorbed L-hydroxyproline (Hyp), investigated alloy positions (one at a time) are indicated by green, Mg atoms are grey. Naming of the alloy positions are atom 63, 88 and 118, left to right. 
Hyp is shown in its relaxed position on 
top of a clean Mg(0001) surface, after relaxation with an alloying atom in one of the three indicated positions all atoms move, both surface and Hyp atoms. 
\textit{Bottom panel:} Relaxed position of Hyp on Mg(0001) with Zn (violet)  in atom position  88. Compared to the starting position
in the top panel Hyp has rotated
to keep one Mg-O binding  and move the O of the hydroxy group loop from on top of the Zn atom to on top of a neighbouring Mg atom of the 
top layer. 
A similar rotation is seen when Zn is in the position of atom 118, thus avoiding O on top of Zn.}
\end{figure}

In all DFT calculations, the atomic positions are optimised using the BFGS algorithm \cite{broyden1970,fletcher1970,goldfarb1970,shanno1970,bfgs}, unless 
explicitly held fixed (such at the lower three layers of Mg atoms).
The convergence threshold for the total energy is $10^{-6}$ atomic units (a.u.)
and for the forces $10^{-5}$ a.u.

We calculate the adsorption energy of the amino acids to the surfaces as the energetic cost of removing 
the molecule from the surface and putting it in isolation,
\begin{equation}
E_\ads = E_\full -E_\mol -E_\surf
\end{equation}
where $E_\ads$ is negative when the molecules bind to the surface,
and $E_\full$, $E_\surf$ and $E_\mol$ are the total energies of the full system, the surface,
and the isolated molecule, respectively.
With this definition a negative energy indicates an exothermic (stable) adsorption.

\begin{figure}[tb]
\centering
\includegraphics[width=0.24\columnwidth]{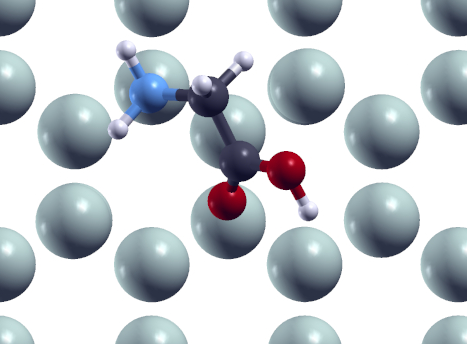}
\includegraphics[width=0.23\columnwidth]{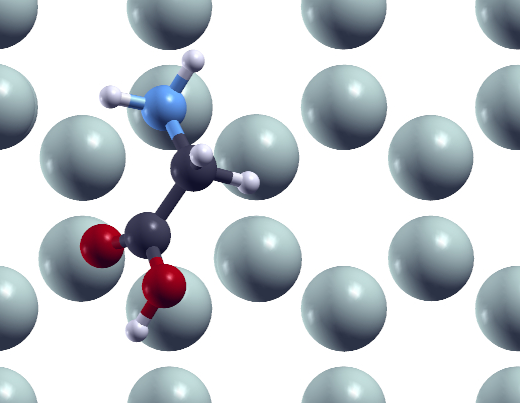}
\includegraphics[width=0.24\columnwidth]{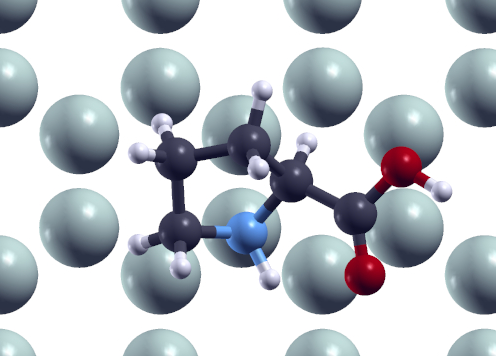}
\includegraphics[width=0.24\columnwidth]{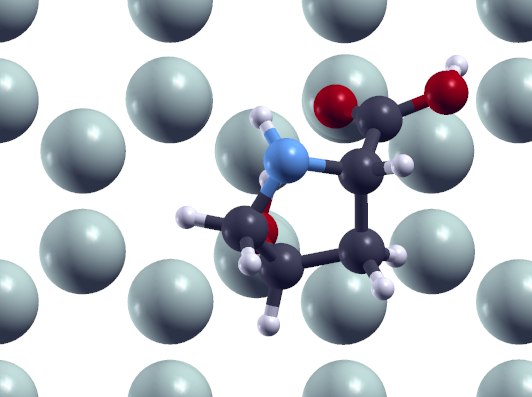}\\[1em]
\includegraphics[width=0.24\columnwidth]{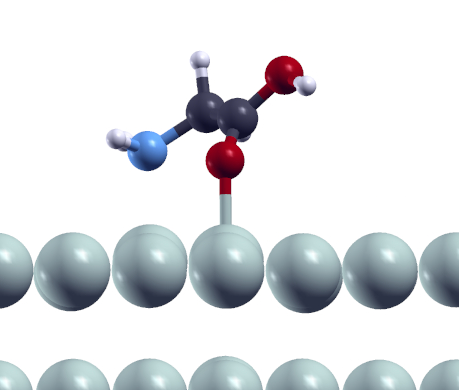}
\includegraphics[width=0.23\columnwidth]{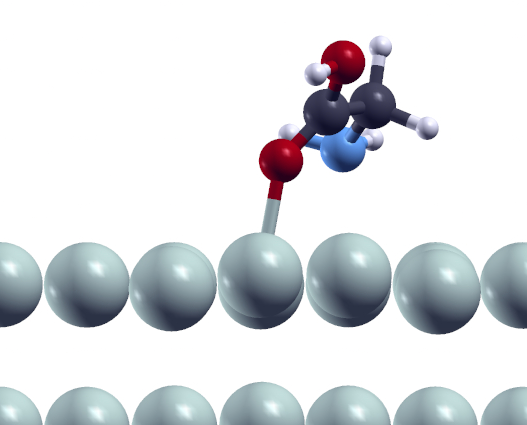}
\includegraphics[width=0.24\columnwidth]{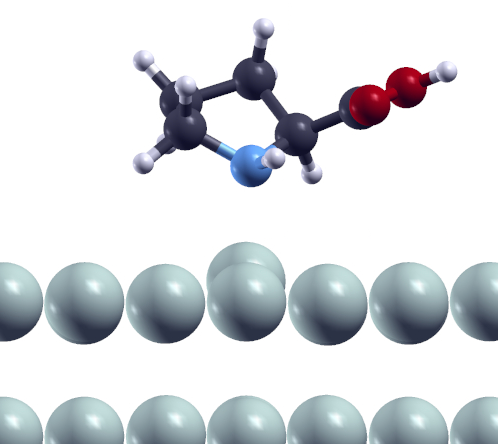}
\includegraphics[width=0.24\columnwidth]{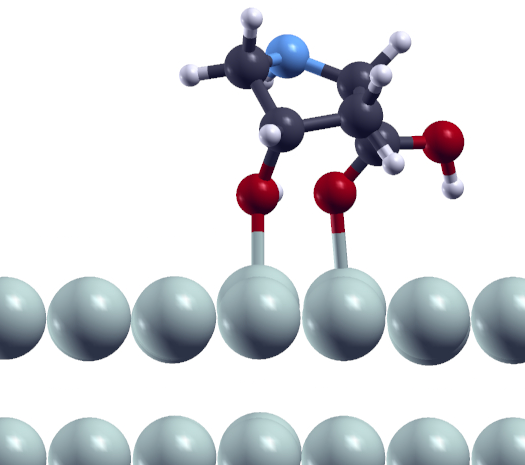}\\
\caption{\label{fig:adsMg}\textit{Left to right:} Glycine in two energetically equivalent positions, L-proline, and L-hydroxyproline, adsorbed on Mg(0001) in vacuum,
top (top panels) and side view (bottom panels). 
Only the central part of Mg(0001) is shown. 
Color coding as in Figures \protect\ref{fig:molecules} and \protect\ref{fig:dopingMg}.}
\end{figure}

\subsection{Alloying considerations}

In doped Mg(0001) we replace a Mg atom in the top layer of the surface with one of Li, Zn or Al.
We only consider doping in top-layer positions because that is where the effect on the molecular adsorption 
is expected to be largest, while second-layer doping 
mainly affects the surrounding Mg atomic positions, and less the molecular binding \cite{fang18}.

We refer to the degree of alloying as the numbers of atom percent (at\%) or weight percent (wt\%) within the two top layers only,
because those are the atoms exposed on the surface.
Replacing one Mg atom in the 50 surface-exposed atoms thus gives the concentration 2 at\%, or 
 0.6 wt\% for Li, 2.2 wt\% for Al, and 5.0 wt\% for Zn.

While less relevant for body implants, we include Al and Li here because they improve the mechanical properties 
of the material and therefore are used in other applications. 
Like Zn, they have Pauling electronegativity (EN) similar to Mg, where EN of an atom is the ability to attract 
electron density, with values in parenthesis
Li (0.98), Mg (1.31), Al (1.61) and Zn (1.65) \cite{pauling60}.
We do not expect changes as profound as from adsorption of (highly electronegative) oxygen atoms 
in Mg surfaces \cite{ScFaKi04,zhe25}.
We do, however, expect that
doping Mg(0001) with Li, Al or Zn changes the electron density around the neighbouring Mg atoms sufficiently to  
affect adsorption of the amino acids. 
We thus expect that the alloying atom or its neighbouring Mg atoms will be more favourable for adsorption of molecular functional groups
than Mg atoms in a clean Mg(0001) surface.

The solubility limits of Zn, Li and Al in Mg, with hcp structure, are 6, 5 and 12 wt\%,
as found from bimetallic phase diagrams \cite{okamoto94,nayeb84,murray82}.
This means that already substituting two Zn atoms in our $p(5\times5)$ surface supercell 
exceeds the solubility limit for Zn in Mg. 

We use Hyp for studying the effect of alloying on the adsorption energy because Hyp contains the most functional 
groups of the studied amino acids. 
We place the alloying atom in one of three
positions relative to the position of Hyp, as adsorbed on clean Mg(0001), and illustrated in Figure \ref{fig:dopingMg}. 
Atom 88 is under the O in the hydroxyl group of the pyrrolidine loop. Atom 118 is placed similarly under the O in 
carbonyl of the carboxyl group at the other end of Hyp. 
While Figure \ref{fig:dopingMg} may indicate that N of the pyrrolidine is above an Mg atom,
it is a second-layer Mg atom and the N atom is positioned several {\AA} from the surface, as also seen in 
the right-most panels of Figure \ref{fig:adsMg}.

To study the effect of alloying atoms next to a given Mg atom, atom 63 (being next to atom 88) was replaced with Zn, Li or Al. 
Atom 63 was chosen as atom 88 (Mg) is directly under the O in the hydroxyl group of the pyrrolidine loop.
After replacing the Mg atom by one of Zn, Li or Al, all atoms (except the three bottom layers of the slab) are again allowed to relax.

\begin{figure}[tb]
\centering
\includegraphics[width=0.4\columnwidth]{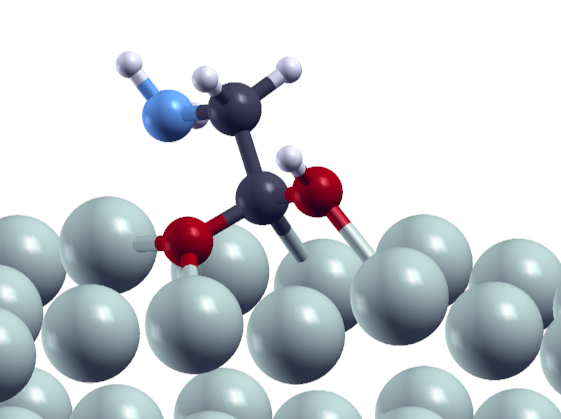}
\hspace{1ex}
\includegraphics[width=0.4\columnwidth]{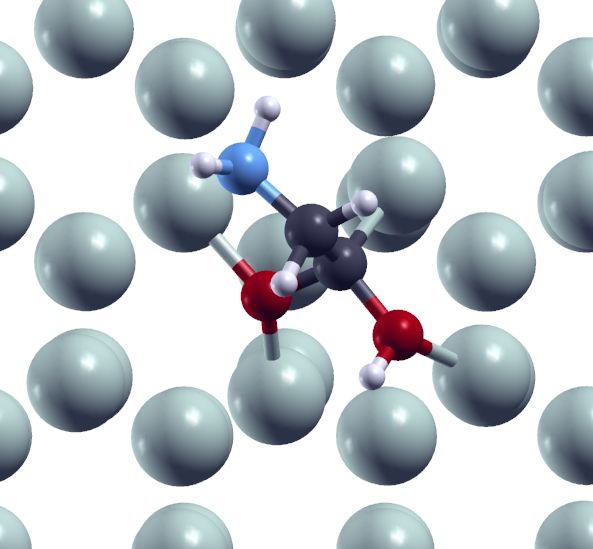}
\\
\caption{\label{fig:adsGlyMg}Glycine adsorbed partly into Mg(0001), side and top view.}
\end{figure}

\subsection{Water environment}

By default, our DFT calculations describe systems in vacuum. 
However, to study the effect of adsorption in a water environment we use the
revised self-consistent continuum solvation (SCCS) model 
of Ref.\ \cite{environ}, as implemented in the Quantum Espresso add-on Environ.
We use a bulk dielectric constant $\epsilon= 78.3 \epsilon_0$ imposed
outside the surface and molecule.  

In the SCCS calculations we keep the periodic boundary conditions in the surface plane, 
but remove the periodicity in the direction perpendicular to the surface. 
We study the system with water  
after further relaxation of the atomic forces in water. 
These further relaxations are usually small, but not always negligible. 

While Mg(0001) will interact with the water molecules where the amino acids do not block access, and 
a realistic body environment also encompasses  other ions, we can at least find an indication of some 
of the changes from our vacuum results when including in our calculations a model of clean water with an effective 
dielectric constant.

The SCCS model of Environ has been tested mainly for a large set of molecules in a series of solutions (including water) \cite{environ,dupont13,hille19}. 
The main quantity used to evaluate the accuracy of the method is the solvation energy, calculated as the difference in total energy $E^\tot$
of the molecule in a water environment and in vacuum
\begin{equation}
\Delta G^\sol = E^\tot_\vac -E^\tot_\wat   \,. 
\label{eq:sol}
\end{equation}
Generally, the errors are less than 1 kcal/mol, but for amines and carboxylic acids the error is reported to be up to 4-5 kcal/mol \cite{environ}.
The molecules of interest for our study thus fall in the less accurate group for solvation energies. 
However, in calculating adsorption energies with these problematic groups present in the adsorbed molecule, 
they are present both in the full system total energy ($E_\full$) with the solution, 
and in the total energy of the molecule isolated in the solution 
($E_\mol$), and there will be at least some error cancellation. 


\section{Results and discussion}
We start by presenting the effect on Mg(0001) of substituting alloying atoms into the surface. 
We then 
calculate and analyse adsorption of the amino acid molecules on clean Mg(0001), 
and of Hyp on Mg(0001) doped with single atoms of Li, Al or Zn. 
Finally, we study how a water-like environment affects the adsorption energies and structures.

\subsection{Mg(0001) surfaces and alloys}

\begin{table*}
\begin{center}
\caption{\label{tab:lattconst}Lattice constants $a$ and $c$ of the 2-atom hcp structure, from Materials Project \cite{materialsproject} (Li, Al, Zn) and 
Ref.\ \cite{schroderPP} (Mg, values for pseudopotential used here),
 here-calculated top-to-second layer separation, $d_{12}$, bond lengths nearest-neighbour Mg to alloying atom X, $d_\MgX$, alloy atom and nearest-neighbour vertical
deviation from first-layer Mg plane, $\Delta h_\alloy$ and $\Delta h_\nn$ (see text and Figure \protect\ref{fig:sketch}), and buckling of the second Mg layer, $\Delta h_2$. 
Upwards changes positive, all values in {\AA}.}
\begin{tabular}{lccccccc}
   & $a$                  &  $c$                 & $d_{12}$              &$d_\MgX$&$\Delta h_\alloy$& $\Delta h_{\nn}$ &$\Delta h_2$\\
\hline
Mg & 3.192$^a$& 5.186$^a$& 2.633$^a$ &3.192, 3.192$^c$ &        0          & 0        &0\\
Li & 3.01$^b$ & 5.11$^b$ & 2.646                 &3.193 & $-0.026$          & $-0.009 $&$+0.024$ \\
Al & 2.81$^b$ & 4.87$^b$ & 2.655                 &3.136, 3.138$^c$& $-0.248$, $-0.246^{c,d}$ & $-0.017$ &$-0.016$\\
Zn & 2.61$^b$ & 4.87$^b$ & 2.650                 &3.116, 3.131$^{c,d}$ & $-0.335$, $-0.273^{c,d}$   & $-0.022$ &$-0.008$\\
\end{tabular}
\end{center}
\footnotesize
$^a${Values for pseudopotential \texttt{Mg\_pbesol\_paw\_3d.UPF} \cite{schroderPP}.}\\
$^b${From Materials Project \cite{materialsproject}.}\\
$^c${DFT calculated values from Ref.\ \cite{zhang21}.}\\
$^d${Values carefully read off from graphs in Ref.\ \cite{zhang21}.}
\end{table*}

\begin{figure}[tb]
\centering
\includegraphics[width=0.99\columnwidth]{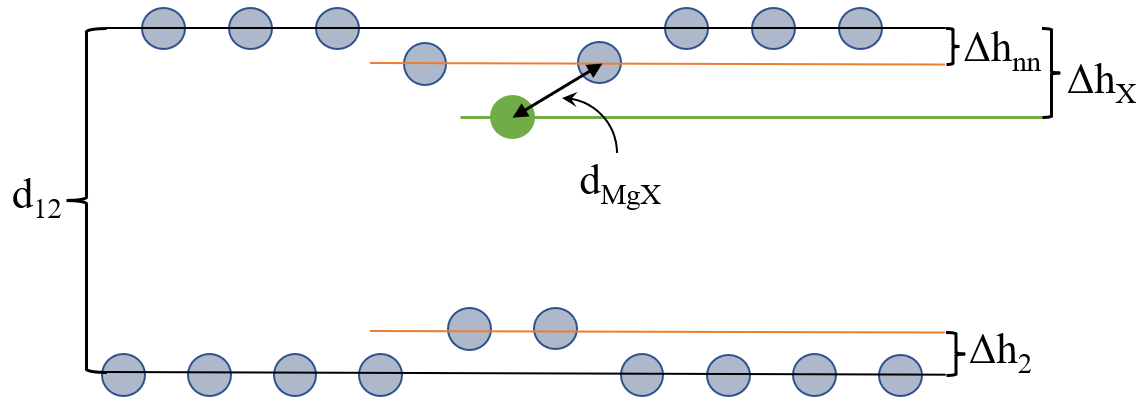}
\caption{\label{fig:sketch}Sketch of the two top layers of Mg (gray) in Mg(0001), with an alloying atom indicated (green), and
exaggerated vertical position changes.
The $\Delta h$ values are negative if the measured atoms are below the plane.}
\end{figure}

The Mg(0001) surface is known to have a slight expansion of the topmost Mg layer separation, compared to bulk; 
for the specific Mg PP used here one of us previously found the expansion $+1.55$\% 
compared to  the bulk layer separation \cite{schroderPP}. 

The bulk hcp lattice constants $a$ and $c$ of Li, Al and Zn \cite{materialsproject} are listed in Table \ref{tab:lattconst}, 
along with the Mg lattice constants found for our choice of exchange-correlation functional and pseudopotential \cite{schroderPP}. 
The hcp $a$ and $c$ values for Li, Al and Zn are smaller than the Mg values by 1.5-18\%. 
We therefore expect some changes in the surface atomic positions once a Mg atom is replaced by an alloying atom, but also that the overall Mg(0001) structure will be kept without any reconstructions of the surface, given the low dosages. 
A sketch used for illustrating the distances and changes is shown in Figure \ref{fig:sketch}, and calculated values are given in Table \ref{tab:lattconst}.

Upon substituting one of the 25 top-layer Mg atoms by an alloy atom, all atoms in the two top layers are allowed to relax.
We determine the heights of the two upper Mg layers and their separation $d_{12}$ from the averages of the vertical positions of Mg atoms in the layer, 
not including the alloying atom and its six Mg nearest neighbours (nn) in the top layer, and the three Mg atoms closest to the alloying atom in the second layer, Fig.\ \ref{fig:sketch}. 
We find that their separation $d_{12}$ expands further, compared to $d_{12}$ of clean Mg(0001), regardless of alloy species, Table \ref{tab:lattconst}. 

For Li, further changes to the structure are modest: the position of Li is $\Delta h_\alloy = -0.026$ {\AA}, under the top Mg plane, 
the six nn Mg atoms are $\Delta h_\nn = -0.009$ {\AA} under the top Mg plane, and the nn Mg-to-Li distance $d_\MgX$ is 3.193 {\AA}, 
only slightly larger than the distance 3.192 {\AA} in clean Mg(0001). 
As we also see for the other two species (below), the second layer is slightly affected, here by buckling upwards directly underneath the Li atom, at $\Delta h_2 = +0.024$ {\AA}. 
Such small changes are reasonable, given that hcp Li has only slightly smaller lattice constants than hcp Mg.

In the surfaces with Al or Zn alloying, the changes are more pronounced, although still keeping the overall Mg(0001) structure. 
The Al (Zn) atom is positioned $\Delta h_\alloy =-0.248$ {\AA} ($-0.335$ {\AA}) lower than the top Mg layer, markedly more so than for Li. 
The distance to the nn Mg atoms is smaller than in clean Mg(0001), with $d_\MgX =3.136$ {\AA} (3.116 {\AA}) for Al (Zn), 
reflecting the smaller in-plane $a$ lattice constants in the Al and Zn hcp phase than in Mg, Table \ref{tab:lattconst}.
The nn Mg atoms move both laterally and vertically. 
The vertical position of the nn Mg atoms is in both cases lower than the top plane. For Zn the top Mg layer buckles down towards the Zn atom, while for Al the next-nearest-neighbour Mg atoms move up, causing small ripples in the plane. 
The second Mg layer in both cases buckles slightly down underneath the alloying atom, $\Delta h_2 < 0$, in contrast to the
upward buckling in the Li case. 
 
With the buckling of the Mg planes and small changes of in-plane positions of the atoms, the local Mg-X distances are generally adjusted to smaller distances than 
the bulk Mg-Mg distances. 
This is also in accordance with the smaller distances between atoms in Al and Zn hcp bulk, Table \ref{tab:lattconst}, 
while Li has lattice constants closer to those of Mg.

Zhang et al.\ \cite{zhang21} studied the changes in (among other properties) atomic structure when single alloying atoms are substituted into the top 
layer of Mg(0001).
They used a $p(4\times4)$ supercell, compared to our $p(5\times5)$, giving a larger concentration of alloying
atoms (1/32) than in the present study (1/50) with one alloy atom in the system, but otherwise their study was carried out similarly to the present study.
They found that all their considered alloying elements (As, Ge, Cd, Zn, Ga, Al), except the large atom Y, 
give rise to a dip in the surface, i.e., a negative $\Delta h_\alloy$.
However, they did not see the small ripples around Al that we see, but with their $p(4\times4)$ supercell 
there are very few third-nearest-neighbour Mg atoms available in the top layer to measure the ripples against.

The distance $d_\MgX$ was also measured by
Zhang et al.\ \cite{zhang21} who found, like here, that the distance decreases compared to the Mg-Mg distance in clean Mg(0001). 
In fact, for Al and Zn our results for $\Delta h_\alloy$ and $d_\MgX$ are in very good (Al) or good (Zn) agreement with those of Ref.\ \cite{zhang21}, see Table~\ref{tab:lattconst}.

\begin{table*}[tb]
\caption{Adsorption energies of the molecules on a $p(5\times5)$ supercell of the Mg(0001) surface.
Hyp ``alloy'' denotes the position of the one substitutional alloying atom per 50 Mg atoms exposed on the surface 
(see indicated atom number in Figure \protect\ref{fig:dopingMg}). Energies in eV/molecule. 
\label{tab:Eads}
}
\begin{center}
\begin{tabular}{l|rrrr|rr}
 & \multicolumn{4}{c|}{Vacuum} & \multicolumn{2}{c}{Water}\\
  &\multicolumn{1}{c}{Mg} & \multicolumn{1}{c}{Li} &  \multicolumn{1}{c}{Al} & \multicolumn{1}{c|}{Zn} & \multicolumn{1}{c}{Mg} &\multicolumn{1}{c}{Zn}\\
\hline
Gly in-surface&       -1.63&       &               &                      & -1.25
\\
Gly on-surface&       -1.05&   &                   &                      & -0.95 
\\
Pro           &       -1.03&     &                 &                      & -0.89 
\\
Hyp, alloy 63 (nn to Mg under pyrr. -OH) & -1.28& -1.29& -1.31                & -1.31                & -0.82  & -0.84
\\
Hyp, alloy 88  (under =O) & -1.28& -1.32& -0.96                & -1.36$^a$& -0.82  & -0.88$^a$  
\\
Hyp, alloy 118 (under pyrrolidine -OH) & -1.28& -1.37& -1.38$^a$& -1.39$^a$& -0.82  & -0.88$^a$ 
\\
Gly-Hyp-Pro   &       -1.79&    &                  &                      & -1.32 
\\
\end{tabular}
\end{center}
\footnotesize
$^a${Molecule rotates such that O above the alloy atom moves to a nearest-neighbour top-layer Mg atom, resulting in two O-Mg bonds to alloy-nearest-neighbour Mg atoms.}
\end{table*}


\subsection{Amino acids on clean Mg(0001)}

\begin{figure}[tb]
\centering
\includegraphics[width=0.5\columnwidth]{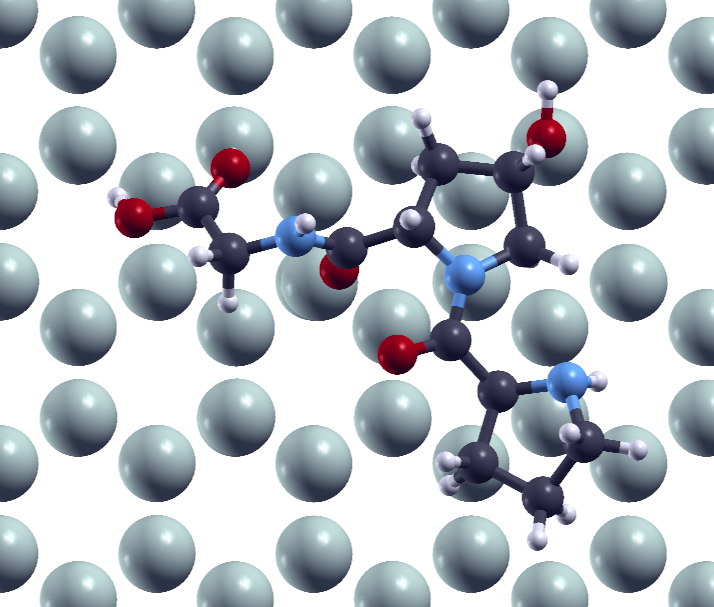}  
\includegraphics[width=0.5\columnwidth]{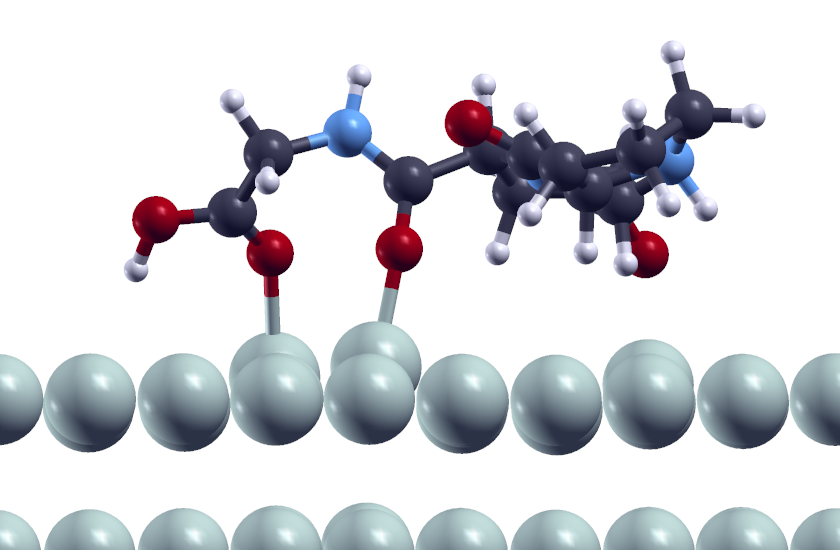}
\caption{\label{fig:mg-glyhyppro}
Mg(0001) with a short snippet of a collagen strand, consisting of Gly-Hyp-Pro. 
Shown are the final positions of the atoms after minimising the forces and thus the energy.}
\end{figure}

When molecules are adsorbed on Mg(0001) both the surface and the molecule adjust their atomic positions. We first focus on
adsorption on clean Mg(0001), i.e., without any alloying atoms in the surface, and calculated in a dry (vacuum) environment.
The adsorption energies, $E_\ads$, of the amino acids and the collagen Gly-Hyp-Pro snippet are compiled in the ``Vacuum'' column ``Mg'' of Table 
\ref{tab:Eads}, and the corresponding adsorption geometries are shown in 
Figures \ref{fig:adsMg}, \ref{fig:adsGlyMg}, and \ref{fig:mg-glyhyppro}.

We note that the adsorption energies lie in the range $-1.6$ to $-1.0$ eV for the single amino acid molecules, while the collagen snippet that contains all three
of them has a slightly stronger bond at $-1.8$ eV, despite being almost three times the size of the individual amino acid molecules.
Each of the amino acids carry a few lone-pair electrons, 
on the O atoms of the carboxyl (-COOH) group and on the N atom of the amino (-NH$_2$) group (Gly) or of the pyrrolidine loop (Pro and Hyp). 
These can potentially bind to metal atoms such as the Mg atoms in the surface.
In the absence of ions from, e.g., H$_2$O in the environment, 
these lone-pair electrons may generate bonds to the Mg atoms of the surface by transfer of electron charge to and from the valence electrons of the 
Mg atom and the O or N atoms.

We find that in all three molecules, the O and/or N atoms form bonds to the Mg surface atoms.
Gly has stable adsorption with both relatively weak and relatively strong 
adsorption energy, corresponding to Figure panels \ref{fig:adsMg}(a) and (b) for the weaker binding and Figure \ref{fig:adsGlyMg} for the stronger bonded situation. 
In the situation with weaker adsorption of Gly (for both of the energetically equivalent structures), 
the =O of {-COOH} has O-Mg bond length 2.147 {\AA} while the  -OH part in -COOH is far from the surface. 
The N in -NH$_2$ has a N-Mg bond length 2.310 {\AA}.
These O-Mg and N-Mg distances for Gly on Mg(0001) are in agreement with the work by Fang et al.\ \cite{fang18}, 
while they find $E_\ads$ at $-1.12$ eV/molecule, slightly stronger than our result $-1.05$ eV/molecule. 
We find that the intra-molecular
structure is not significantly affected by the adsorption. 

Meanwhile, we find that the Gly adsorption structure with a stronger adsorption energy (Fig.\ \ref{fig:adsGlyMg}) has a more complex adsorption. 
Gly has both O atoms of -COOH forming O-Mg bonds. Four Gly atoms are close to the Mg surface: 
The =O atom is at distance 1.944 {\AA} and 2.009 {\AA} from two neighbouring Mg atoms, the O of -OH 
has the distance 2.159 {\AA} to another Mg atom, the N atom of -NH$_2$ forms a bond at length 2.280 {\AA} to the Mg atom shared with =O, 
and finally the C atom of -COOH is also dragged close the surface with distance 2.216 {\AA} to a Mg atom.
This stretches the intramolecular O-C distance to 1.4 {\AA} from the value 1.2 {\AA}
both in the isolated weakly bonded Gly.
The =O atom is adsorbed into Mg(0001) in a position resembling a tetrahedral site of single 
O adsorption \cite{ScFaKi04,FrTa13,xing24} while the remaining atoms
stay in positions on top of the surface.
With an O atom into the surface and further atoms close to the surface,  this structure naturally has a larger adsorption energy, at 
$E_\ads=-1.63$ eV compared to the weaker $E_\ads=-1.05$ eV with one O-Mg and one N-Mg bond.

In Pro adsorption, the N atom of the pyrrolidine loop consistently approaches the Mg surface in our relaxations, independently of the various starting 
situations studied. 
The N-Mg distance is 2.278 {\AA}, while the two O atoms of -COOH are much further from the surface at 3.673 {\AA} (=O) and 4.223 {\AA} (-OH). 
In this adsorption geometry the Mg atom under N is dragged out about 0.4 {\AA} from the average height of the top Mg layer, while the Mg atom under -OH sinks 
into the surface by 0.1 {\AA}. 
It is therefore clear that steric hindrance is not the reason for the binding to N and not to the O atoms.

While Hyp has a similar structure as Pro, the added -OH group on the pyrrolidine loop becomes important in the adsorption, at O-Mg distance 2.139 {\AA}. 
At the same time, the =O of -COOH is also close to the surface, at 2.198 {\AA}, while N is not near any Mg atom. 
In other words, the added -OH group of Hyp compared to Pro makes a large difference in the adsorption structure. 

The O-Mg and N-Mg distances found above for (weakly bound) Gly, Pro and Hyp correspond to the distances found for the same functional groups 
in other amino acids and dipeptides in Ref.\ \cite{fang18}.
In the discussion above and below, one should keep in mind that also the parts of 
the molecule at longer distances from the surface interact with the surface through, e.g., vdW interactions, 
but contributing a smaller part of the adsorption energy.

For a more realistic model of collagen we study a three-amino-acid snippet of a collagen strand, Figure \ref{fig:mg-glyhyppro}. 
The difference from the adsorption of the 
amino acids one by one is that several of the reactive groups in the adsorption of the single amino acids are now used to bind the amino 
acids into the collagen snippet. 
The adsorption energy is therefore not the sum of the amino acid values, Table \ref{tab:Eads}, and
is dominated by interactions between =O and Mg atoms, as seen in the side view of Fig.\ \ref{fig:mg-glyhyppro}.

\subsection{Adsorption on alloy surfaces}

To learn how the adsorption is affected by sparse alloying in the Mg surface, 
we here focus on the adsorption of Hyp.  
Hyp has both a carboxyl group (-COOH), an amine group (-NH) in a pyrrolidine ring (in lieu of the 
amino group, -NH$_2$ of most other amino acids), and a hydroxyl group (-OH) attached to a C atom in this ring. 
This creates four potentially reactive parts of the molecule. 
We find from the adsorption of Hyp on clean Mg(0001) that only two of these groups end up close to any of the surface atoms.
We study the effect of exchanging a Mg atom by an alloy atom (Li, Al, Zn) under or near the functional groups that are close to the surface in
clean Mg(0001), as explained in the methods section.

The relaxed position of Hyp on clean Mg(0001), right hand panels of Figure \ref{fig:adsMg},
is the starting position. 
There, two O atoms are close
to Mg atoms in the surface,
whereas the N atom is further from the Mg atoms.
We calculate the adsorption energy at the alloyed surface and analyse the changes in positions of the atoms, 
with alloying atoms in one of the three positions atom 63, 88 and 118, see top panel of Figure \ref{fig:dopingMg}.

From our nine calculations of Li, Al, or Zn in one of these three positions we consistently find that the O atoms of Hyp 
are attracted to Li but are repelled from Zn and Al, as expected from the EN values.
The repulsion either results in longer (and thus weaker) O-alloy bond lengths or in 
the O atom moving to the nearest Mg top-layer atom, by rotating the molecule, Figure \ref{fig:dopingMg} bottom panel. 
The attraction to Li results in shorter alloy-O distances when O is above Li, i.e., with Li in positions 88 or 118. 

Focusing on the atom distance changes of Mg or alloy atom to O, of O in pyrrolidine -OH and =O in carboxyl, we find that 
with Li in a neighbouring position (atom 63) the structure of the adsorbed molecule relative to the surface hardly changes, 
compared to adsorption without alloys, with the
Mg-O distances changing less than 0.006 {\AA}.
As in the Mg(0001) surface with Li and without molecules, the Li atom moves slightly into the surface.
With Li under one of the O atoms, the changes are remarkably different: the distances decrease by about 0.2 {\AA} above the Li atom
and 0.03-0.04 {\AA} for the Mg-O distance for the other of the two binding O atoms. This is reflected in the almost unchanged binding energy for
the situation with 'atoms 63' while the adsorption energy improves by 0.04 to 0.09 eV/molecule for the situations with Li under an O atom, Table \ref{tab:Eads},
column 'Vacuum' with 'Li'.

With Al or Zn as the alloying atom, the repulsion is seen in situation 'atom 88' and 'atom 118' either as a longer alloy-O distance 
(in Al 'atom 88' the length increases
by 0.06 {\AA} with a major loss of adsorption energy 0.32 eV/molecule) 
or a rotation of Hyp to fit the two O atoms on top of Mg atoms that are nn to the alloying atom. When rotated or already with the alloy
as a nn ('atom 63'), the distances either decrease by up to 0.02 {\AA} or remain largely unchanged.
The one exception is Zn in position 'atom 88' where Hyp does rotate, but the two Mg-O bond lengths both increase (for -OH, at 0.04 {\AA}) 
and decrease considerably (for =O, at 0.09 {\AA}). In all structures with O atoms 
above Mg atoms (either from the start, as in 'atom 63', or after rotation of Hyp) the adsorption energy increases, by 0.03 eV/molecule when only one of the 
Mg atoms are nn to the alloying atom (in 'atom 63' situation) or by 0.08-0.11 eV/molecule when both Mg atoms are nn to the alloying atom. 
We therefore conclude that the Mg atoms that are nn to alloy atoms do indeed have their electron structure changed to become more favourable for binding 
than Mg atoms further away. 

Overall, alloying improves the adsorption energy, Table \ref{tab:Eads}. Although the 'atom 88' position for Al gives rise to a worse 
adsorption energy than without the alloy, the other two positions are more favourable and it is therefore not likely that
the molecule would end up in that particular orientation in adsorption on a surface that already has alloying atoms.

\begin{table}[tb]
\caption{Solvation energies of the molecules, fully relaxed in solution and with the geometry of vacuum. 
Calculated in the $p(5\times5)$ hcp unit cell of height 36 {\AA}. Calculated energies in eV/molecule (first number) and kcal/mol (second number). 
Reference values in kcal/mol. 
\label{tab:Esolv}
}
\begin{center}
\begin{tabular}{l|cc|c}
 & \multicolumn{1}{c}{$\Delta G^\sol$ } & \multicolumn{1}{c|}{$\Delta G^\sol_\vg$}& \multicolumn{1}{c}{ref.} \\
\hline
Bz  &-0.033/-0.76& -0.032/-0.75& /-0.87$^a$ \\
Gly &-0.45/-10.3& -0.42/-9.8& /-10.7$^b$\\
Pro & -0.34/-7.9& -0.33/-7.6& \\
Hyp & -0.58/-13.3& -0.56/-12.9&\\
\end{tabular}
\end{center}
\footnotesize
$^a${Minnesota Solvation Database \cite{minnesota}.}\\
$^b${Ab-initio HF+MP2 calculation (def2-TZVP with 185 basis functions) \cite{boca23}.}
\end{table}

\subsection{In a water environment}

We first evaluate the solvation energies of Gly, Pro and Hyp in water within the model provided by Environ. 
We are not a priori interested
in the solvation energies, but we do determine the solvation energies of our molecules in the unit cell used in all of our calculations 
(now without the Mg atoms), to see how much the periodicity (in two directions) and the relatively narrow side lengths (compared to the 
original solvation energy calculations) affect the energies.
We also separate out the part of the solvation energy $\Delta G^\sol_\vg$
that only relates to the difference in 
environment by calculating the solvation energy based on the total energy in water using the geometry of the
molecule as optimised in vacuum in both right-hand terms of (\ref{eq:sol}). We include here also calculations of benzene (Bz) 
in order to compare to literature values, Table \ref{tab:Esolv}. 

The solvation energy results show (for Bz and Gly in the solution-geometry, first column) reasonable agreement with literature values (right hand column).
For Pro and Hyp we were not able to find corresponding literature values. 
We also find that the geometry optimization in water has a relatively small but significant contribution to the solvation energy, at around 
0.02 eV/molecule.
In all further calculations with water we included relaxations.

The right hand part of Table \ref{tab:Eads} lists $E_\ads$ of Gly, Pro, Hyp and the Gly-Hyp-Pro snippet on 
clean Mg(0001) when in the modelled water environment. The table also lists $E_\ads$ for Hyp on Mg(0001) with a Zn atom in the previously discussed positions.

With the exception of Pro and Gly adsorbed \textit{on\/} the surface, all adsorption energies lose 0.4-0.5 eV/molecule by being embedded in the 
water-like environment. 
All these adsorbed molecules have in common that at least two O atoms from -COOH and/or -OH of the pyrrolidine loop (Hyp) form close 
bonds to the surface. 
In Pro and Gly on the surface, N-Mg bonds instead dominate, Gly having only one O-Mg bond, 
and for these molecules the decrease in adsorption energy is only 0.1 eV. 

The dielectric constant outside of the surface and molecules reduces the charge relocalisation relative to vacuum conditions. 
In vacuum conditions, a single N atoms adsorption energy is around 1/4th that of a single O atom, resulting in less charge relocation. 
Therefore, the Mg-N bonds in the molecules discussed here are less affected by the dielectric constant than the Mg-O bonds. 
The effect of this is seen in Table \ref{tab:Eads}.

A closer look at the geometry changes in water shows that in water the bond lengths are changed, mainly decreasing approximately 0.02 {\AA} for Pro and Hyp,
except the Mg-O distance of =O in Hyp which instead increases by 0.02-0.03 {\AA} on clean Mg(0001) and with Zn in positions 'atom 63' and 'atom 118'.
For Gly on the surface changes are negligible, and for Gly in the surface all bonds with the surface decrease  by 0.01 to 0.07 {\AA}.
In other words, there is some variation in the structure, but all adsorption energies become less strong in this model of water, already before including 
dissociated water molecules and other ions.

\section{Conclusion}
With density functional theory calculations we studied pure and alloyed magnesium surfaces and their interaction with the amino acids Gly, Pro and Hyp,
as well as with a collagen snippet consisting of these three 
amino acids.
The background for our study is the wish to better understand how magnesium-based degradable implants can be improved and suitably protected.
The amino acids may be seen as part of the collagen of the body in the surrounding of the implant, 
but also as potential coating of the magnesium surfaces to inhibit corrosion.

We find good adhesion of the amino acids on the surface, with increased adsorption strength for sparse  alloying by Li, Al or Zn.
In an environment with water represented by its static dielectric constant the adsorption energies 
become less strong, while the molecules are almost all brought closer to the surfaces, with shorter bonds between atoms in the surface and in the molecules. 
However, the molecules are also in this situation found to bind to the surface.

We present the changes in atomic positions of Mg(0001) when the surface is sparsely alloyed (by 2 at\% among the exposed Mg atoms of the surface), 
and the adsorption energies and bond length changes when amino acids are adsorbed on these surfaces and on pristine Mg(0001), under dry and water conditions.

We note that the studied Mg(0001) is the least reactive of the surface orientations of Mg \cite{xing24}, in future studies it
would therefore be of interest to study also others of the common Mg surfaces, as well as surfaces with defects. These may also
be relevant because corrosion is mostly
found in cracks and pits of the material, deformations in which several surface orientations and defects will be present. 

\section*{acknowledgement}
We thank Dmytro Orlov, Lund University, and Petra Maier, University of Applied Sciences, Stralsund, for helpful discussions and feedback.
J.B., A.G., O.H., and A.L.\ contributed equally to this work. 


The present work is supported by the Swedish Research Council (VR) through Grant No.\ 2020-04997, 
the Swedish Foundation for Strategic research (SSF) through Grant IMF17-0324, 
and by Chalmers Area of Advance Nano.
The computations were performed using computational and storage resources at 
Chalmers Centre for Computational Science and Engineering (C3SE), 
and with computer and storage allocations from the 
National Academic Infrastructure for Supercomputing in Sweden (NAISS), under contracts
NAISS2024/3-16, 
NAISS2024/6-432, 
and NAISS2025/3-25.

The authors declare no conflicts of interest.

%

\end{document}